\documentclass[journal]{IEEEtran}
\usepackage{blindtext}
\usepackage{graphicx}
\usepackage{balance}

\begin{document}

\title{A Low-Power 20-Gb/s Discrete-Time Analog Front-End for ADC-Based Serial Link Equalizers}

\author{ Mostafa M. Ayesh$^{1}$,
Sameh Ibrahim $^{1}$ and Mohamed M. Aboudina $^{2}$
\\
$^{1}$ \IEEEauthorblockN{EECE, Ain Shams University} 
$^{2}$ EECE, Cairo University} 

\maketitle
\begin{abstract}

This paper presents a discrete-time analog front-end for an analog-to-digital (ADC) based equalizers. The front-end uses a discrete-time linear equalizer (DTLE) and ultra-low-power 4-bit time-interleaved charge-steering flash ADC. The DTLE serves two functions; linear equalization and sampling and holding for the following charge-steering ADC. The ADC uses fully differential low-power clocked comparators. Low power in the comparators is achieved by embedding a dynamic latch into the core of a charge-steering pre-amplifier. The 20-Gb/s front-end is designed and simulated in a 65-nm CMOS technology. The flash ADC uses 4-stage interleaving and thus requires 4 DTLEs running at 5 Gb/s. A 5-Gb/s DTLE consumes 0.57 mW from a 1.2-V supply and the ADC consumes 15.5 mW from a 1-V supply at 20 GS/s for a total power dissipation of 17.78 mW or 0.89 pJ/bit. The ADC has an SNDR of 23.9 dB, an SFDR of 33.6 dB, and an effective number of bits (ENOB) of 3.67 bits for a sinusoidal input of frequency 9.84 GHz and amplitude 600 mV $_{diff}$.
\vspace{1ex}

\keywords

Charge-Steering, Preamplifier, Regeneration, Ultra-low power, Comparator, Flash ADC, High-speed ADC, Time-interleaved, ADC-based systems, Dynamic latch, Discrete-Time, Linear Equalizer.
\end{abstract}

\IEEEpeerreviewmaketitle

\section{Introduction}

High-speed comparators with low-power consumption and small sensitivity are essential blocks in many applications that require very high sampling rates and low resolution such as data-storage read channels and wired communication systems [\ref{paper:1}]. As the demand for high rates increases, comparators must cope with this need while achieving as low power consumption and small area as possible. For example, recent communication links try to use ADC-based receivers which mainly depend on analog-to-digital converters (ADCs). However, power consumption remains a concern. A comparator is the main building block in any ADC. Improving its performance by reducing its kick-back noise and power consumption while maintaining high sampling rate will directly improve the whole ADC. As a result, the multi-standard reconfigurable systems which heavily depend on an ADC near the front end and digital blocks at the back end are becoming more common.

\vspace{1ex}

%Designing high-speed comparators in deep sub-micron CMOS technologies with these specifications is challenging. They suffer from low supply voltages especially with the fact that threshold voltages of the devices have not scaled down by the same factor as the supply voltages of modern CMOS processes [\ref{ICM_paper_2}]. Add to this, lowering the supply voltage results in a tight and limited range of input common-mode voltage which is important in many high-speed ADC architectures, such as flash ADCs and in power management blocks, such as buck converters. It also hinders getting higher speeds. %%Many attempts to achieve high-speed comparators with low-voltage supply are available in the literature [\ref{ICM_paper_3} - \ref{ICM_paper_5}]. The comparator in [\ref{ICM_paper_3}] has some modifications in its timing and some additional devices to make it able to work down to 0.5 V on a 600-MHz clock and consumes 18 $\mu$W. However, transistors mismatch in the additional circuitry should be examined.
%In [\ref{ICM_paper_5}], the comparator was expanded into two paths between $V_{DD}$ and ground so that only one threshold voltage of one transistor instead of two exists in each path. These attempts add additional circuits to the StrongARM comparator [\ref{paper:strongArm_origin}] to enhance its speed at low supply voltages.

\vspace{1ex}

\begin{figure*}[!h]
\centering
\includegraphics[height=2in]{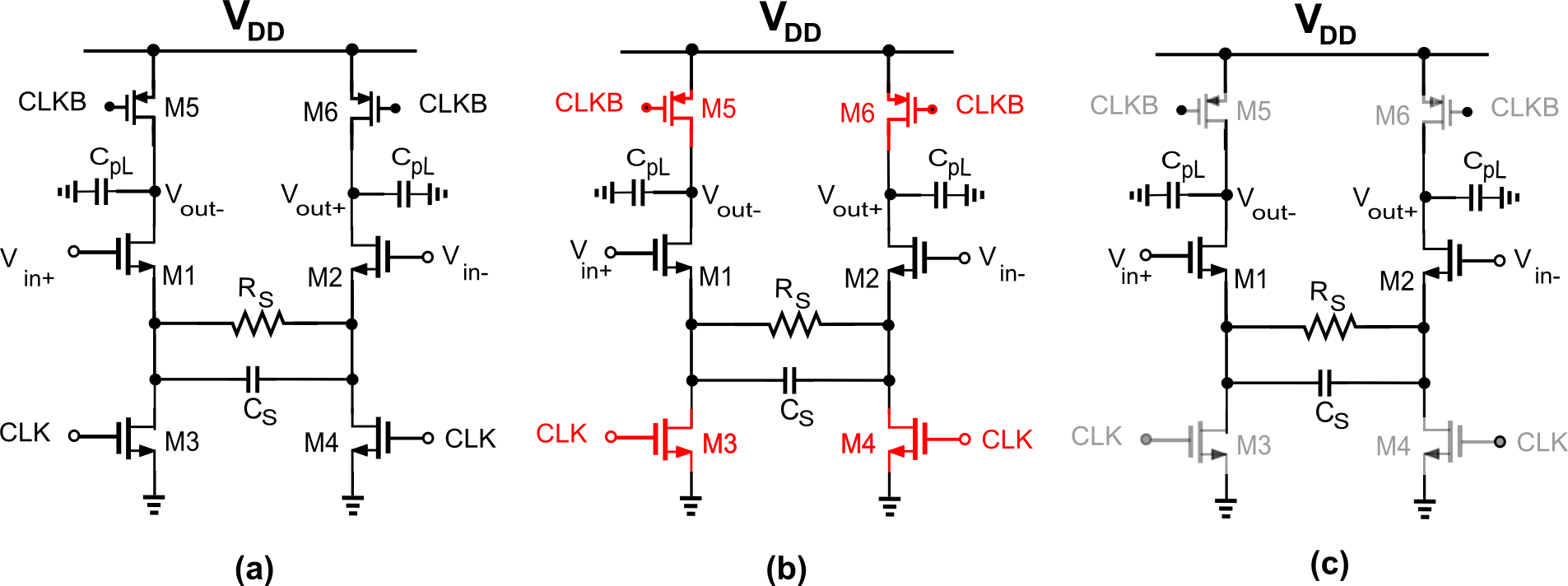}
\caption{(a) A Schematic of the DTLE Circuit. (b) The Track-and-Equalize Mode (CLK = ``High''). (c) The Hold Mode (CLK = ``Low'').}
\label{fig:dtle_sche}
\end{figure*}

\vspace{2ex}
Time-interleaving has been widely used to allow for higher speed ADCs while keeping power consumption at reasonable levels. The power consumption overhead of interleaving becomes insignificant above certain frequencies when compared to the increase of the power consumption of a single-channel ADC.

\vspace{2ex}

The goal of the paper is to propose and analyze a novel topology, based on the charge-steering concept, for an ultra-low-power ultra-high-speed comparator. Then this comparator is used to build a low-power flash ADC to be used in a high-speed serial link ADC-based equalizer. A 4-bit 20-GS/s 4x time-interleaved flash ADC is used. Usually, serial link equalizers need some linear equalization to compensate for precursor inter-symbol interference (ISI), as a result, the design of the analog front-end of the equalizer is completed by adding discrete-time linear equalizer (DTLE) at the input.
%The proposed comparator adopt charge steering concept, it merges the pre-amplifier and the regenerative latch into one block as in [\ref{paper:embedded_latch}] while relying on charge redistribution between different phases of operation to achieve pre-amplification and regeneration. This comparator is proposed without boosting voltage techniques, it also uses the same number, or even less, of stacked transistors found in the conventional dynamic comparators. An analysis about the delay of the proposed dynamic comparator is presented. This analysis can be added or compared to the one in [\ref{ICM_paper_9}] for other various dynamic-comparators topologies.

\vspace{2ex}

The design of the DTLE is explained in section \ref{sec:dtle}. Section \ref{cs_paradigm} discusses the charge-steering concept. The proposed fully differential comparator is discussed in section \ref{sec:CS_COMP} along with its proposed single-ended version and its delay analysis. Section \ref{sec:flash_adc} shows the designed charge-steering Flash ADC based on that comparator. The simulation results of the comparator, the ADC, and its front-end are given in section \ref{sec:sims}. Finally, the work is concluded in section \ref{sec:conc}.

%%%%%%%%%%%%%%%%%%%%%%%%%%%%%%%%%%%%%%%%%%%%%%%%%%%%%%%%%%%%%%%%%%%%%%%%%%%
%%%%%%%%%%%%%%%%%%%%%%%%%%%%%%%%%%%%%%%%%%%%%%%%%%%%%%%%%%

\section{Discrete-Time Linear Equalizer}
\label{sec:dtle}

\begin{figure}[!b]
\centering
\includegraphics[width=3in]{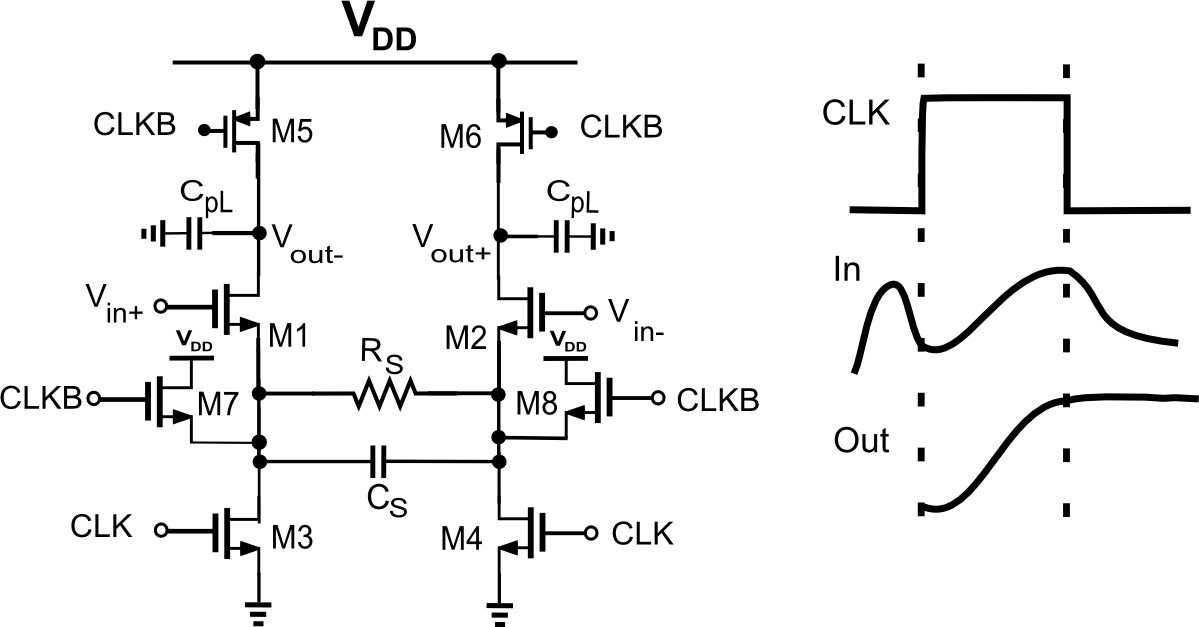}
\caption{The DTLE with Leakage Blockers.}
\label{fig:modified_dtle}
\end{figure}

Since the analog input is changing at very high speeds, Flash ADC-comparators might samples the input at different instances due to the different types of mismatches. To overcome this issue, a sample-and-hold (S/H) circuit is needed at the input of the ADC to make sure that the introduced input to all of the comparators is the same. During the reset mode of the comparators, the sample-and-hold circuit samples the input then hold it for the comparators during the amplification and the regeneration modes [\ref{paper:Ismail_DTLE}].

\vspace{2ex}

For pre-cursor ISI equalization, many designs adopted the conventional continuous-time linear equalizer (CTLE) [\ref{paper:Gondi}]. The main disadvantage of the CTLE is the direct and continuous path from supply to ground. That is why it is not a proper option for the low-power applications. A DTLE is a linear equalizer that replaces the conventional CTLE and its continuous-time nature to a discrete-time nature [\ref{paper:Ismail_DTLE} - \ref{paper:Gondi}]. It could be seen as a combined continuous-time linear equalizer and a sample-and-hold circuit. 

\vspace{2ex}

The resistive loads in a conventional CTLE are replaced with switched resistors and the conventional current sources are replaced with clocked current sources as shown in Figure \ref{fig:dtle_sche}. The circuit has two main modes of operation, the hold mode, and the track-and-equalize mode. When CLK signal is ``High'', the track-and-equalize mode is activated. The clocked current sources and the loading resistors are ``ON'', hence the DTLE works as a peaking amplifier as shown in Figure \ref{fig:dtle_sche} (b). When CLK signal goes ``Low'', the circuit enters the hold mode. The clocked current sources are ``OFF'' and the switched resistors are at high impedance and hence the output values are almost not changing during this phase as presented in Figure \ref{fig:dtle_sche} (c). This is good for the ADC to follow and cancels the need for a sample-and-hold circuit. As a result, power is saved.

\vspace{2ex}
However, the input devices are not instantly switched off because some time is needed to charge the tails nodes to switch off the input devices, so the output nodes face some leakage and the amplitude drops a little. Some reset NMOS transistors $M7$ and $M8$ might be added to rapidly charge the tail nodes and thus switch off the input devices during the hold mode. This will decrease the leakage of the output node and solve the issue as shown in Figure \ref{fig:modified_dtle}. The input at the end of the clock is sampled, processed and held. 

\vspace{2ex}
The circuit is degenerated with a capacitor $C_{s}$ which is the main reason for the gain peaking. This capacitor with the degeneration resistor $R_s$ form a zero at $\omega_{z}$ and a pole $\omega_{p1}$, and these are added to the existing pole at the output node at $\omega_{p2}$. The locations of the poles and the zero are as following: 
\begin{equation} 
\omega_{z} =\frac{1}{R_s C_s} 
\end{equation} 

\vspace{2ex}

\begin{equation} 
\omega_{p1} =\frac{1+(g_m+g_{mb})R_s/2}{R_s C_s} 
\end{equation} 

\vspace{2ex}

\begin{equation} 
\omega_{p2} =\frac{1}{R_d C_{pL}} 
\end{equation} 

\vspace{1ex}
where $R_d$ is the resistance of the clocked PMOS switch in the load that is kept in the linear region of operation.

\begin{equation} 
R_d =\frac{1}{\mu_p C_{ox} (W/L)  (|V_{GS}|-|V_{th}|)} 
\end{equation} 

Some of the most important and meaningful quantities for the peaking amplifier is the DC gain, AC gain and the peaking it achieves. Peaking is defined as the ratio of the high-frequency gain to the low-frequency gain.

\vspace{2ex}

\begin{equation} 
Gain_{DC} =\frac{(g_m+g_{mb})R_d}{1+(g_m+g_{mb})R_s/2}
\end{equation} 

\vspace{2ex}

\begin{equation} 
Gain_{Hi-freq} =g_m R_d
\end{equation} 
\vspace{1ex}

%%%%%%%%%%%%%%%%%%%%%%%%%%%%%%%%%%%%%%%%%%%%%%%%%%%%%%%%%%%%%%%%%%%%%%%%%%%%%%%%%%%%%%%%%%%%%%%%%%%%%%%%%%%%%%%%%%%%%%%%%%%%%%%%%%%%%%%%%%
\section{Charge-Steering Concept} 
\label{cs_paradigm}

\begin{figure}[!h]
\centering
\includegraphics[height= 2in, width=3.4in]{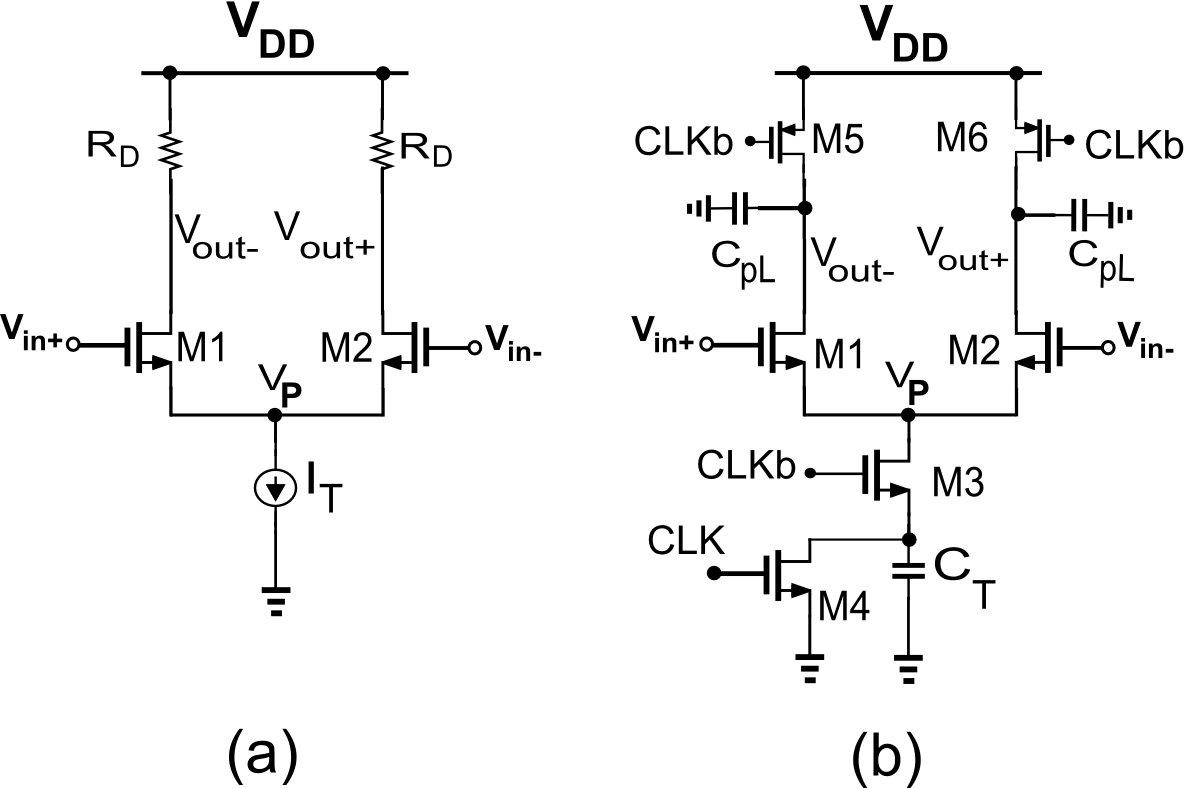}
\caption{Differential Pair Using: (a) Current-Steering. (b) Charge-Steering. [\ref{paper:CS_razavi}] }
\label{fig:current_charge}
\end{figure}

Instead of using current-steering power-hungry circuits, discrete-time charge-steering circuits consume less power than their continuous-time current-steering counterparts, especially at high speeds. This advantage can be utilized in designing semi-analog circuits such as latches, demultiplexers, clock-and-data recovery (CDR) circuits, and comparators as well as mixed-mode systems such
as ADCs [\ref{paper:CS_razavi}]. The main concept is converting the continuous-time current-steering circuit to a discrete-time charge-steering topology as in [\ref{paper:CS_razavi}]. The tail current source, as in Figure \ref{fig:current_charge} (a), is replaced with a charge source, and the load resistors with switched capacitors. Discrete-time operation requires two switches in the tail path and two at the output nodes as shown in Figure \ref{fig:current_charge} (b).

\vspace{2ex}

\begin{figure}[!h]
\centering
\includegraphics[height=2in]{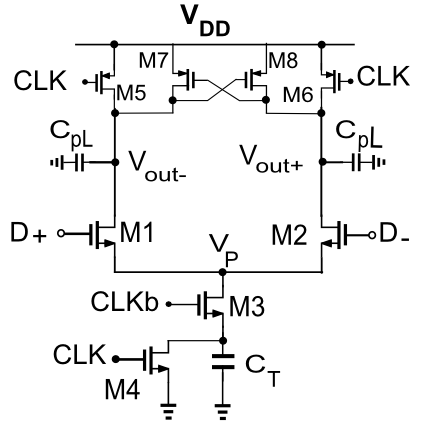}
\caption{A Schematic Diagram for a Charge-Steering Latch. [\ref{paper:CS_Feb2015}]} % - \ref{paper:ayesh_ICM}]}
\label{fig:proposed_comp}
\end{figure}

The operation of this charge-steering-based differential amplifier is divided into two phases; reset phase and amplification phase. In the reset phase, the tail capacitor is discharged while the output nodes are pre-charged to $V_{DD}$. In the amplification phase, the tail capacitor is connected to the tail node, drawing a current from the input pair while the outputs are disconnected from $V_{DD}$. Then, the input pair draws a differential current from the load capacitors in proportion to the differential input voltage. This results in an output voltage proportional to the input voltage. Figure \ref{fig:proposed_comp} was introduced in [\ref{paper:CS_Feb2015}] as a dynamic latch for digital signals utilizing charge-steering circuits. The extra back-to-back PMOS pair $M7$-$M8$ helps in extending the swing of the original amplifier. This circuit cannot be used as a comparator for analog signals. The reason is that for small input signals, the input pair $M1$-$M2$ is fighting against the regenerative latch $M7$-$M8$ and will need a long time constant to produce an output. Hence, at high clock frequencies, the sensitivity of this circuit is compromised. %To overcome that effect, larger input devices, as well as larger bottom capacitors, must be used increasing the circuit power consumption. A solution to this issue is proposed in [\ref{paper:ayesh_ICM}] where a different timing scheme is utilized. 

\section{The Proposed Fully-Differential Comparator} 
\label{sec:CS_COMP}

\begin{figure}[!h]
\centering
\includegraphics[width=3in]{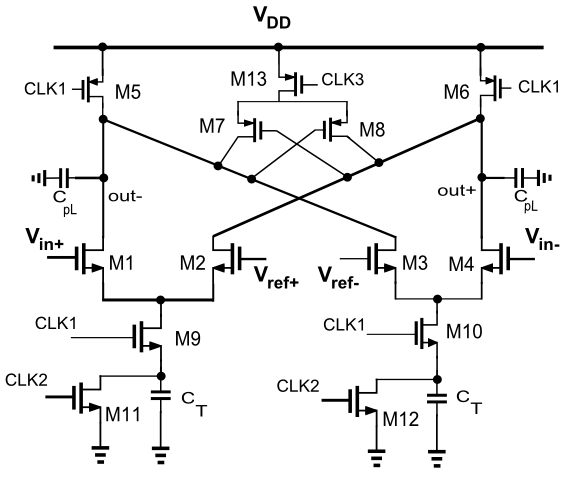}
\caption{Proposed Fully-Differential Latched Dynamic Charge-Steering Pre-Amplifier. [\ref{ICM_paper_15}]}
\label{fig:proposed_comp_diff}
\end{figure}

\begin{figure}[!h]
\centering
\includegraphics[height=1in, width=1.5in]{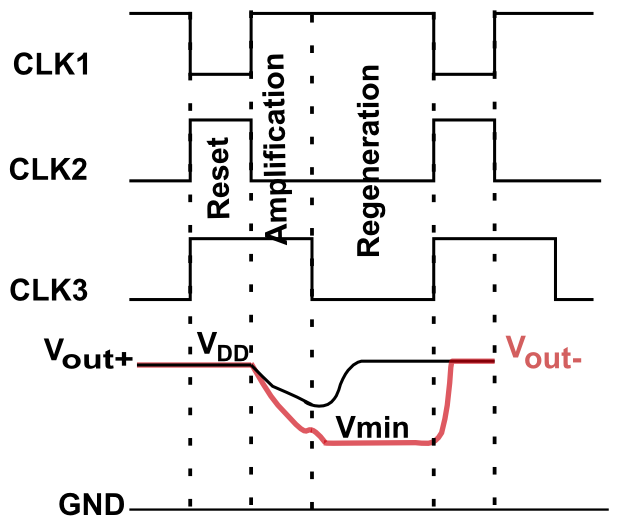}
\caption{Clocking Scheme for Different Phases.}
\label{fig:time}
\end{figure}

\begin{figure*}[!h]
\centering
\includegraphics[width = 7in]{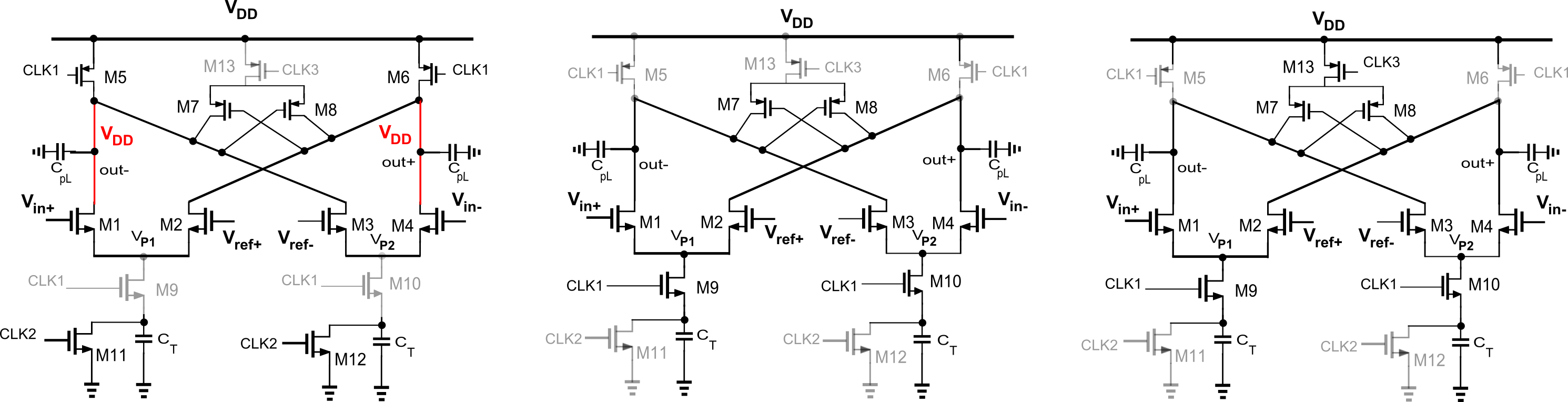}
\caption{The Proposed Comparator (a) Reset Phase. (b) Amplification Phase. (c) Regeneration Phase.}
\label{fig:circuit_phases}
\end{figure*}

The proposed comparator is utilizing the charge-steering concept as shown in Figure \ref{fig:proposed_comp_diff}. The basic charge-steering amplifier consists of transistors $M1$-$M6$, $M9$-$M12$ and two tail capacitors $C_T$ amplifies the input signal. In order to regenerate the output signal, a regenerative latch might be added after this stage which will cause an increase in the power consumption of the comparator drastically. In this proposed design, an idea of an embedded regenerative latch as in [\ref{paper:embedded_latch}] is used but it is turned ``OFF'' in the reset and amplification phases and is turned ``ON'' in the regeneration phase. This is achieved by switching the transistor $M13$ in Figure \ref{fig:proposed_comp_diff}. $M13$ isolates the regenerative latch from both reset and amplification phases in order to avoid fighting. This results in an ultra-fast amplification for very low power consumption. The role of the embedded regenerative latch is to redistribute the differential charge increasing the output voltage. A new clocking scheme is introduced to facilitate the proposed operation. Figure \ref{fig:time} shows the proposed clocking scheme. The operation is divided into three phases of operation: reset phase, amplification phase, and regeneration phase. Figure \ref{fig:circuit_phases} shows the proposed circuit in its three phases of operation.

\subsection{Reset phase}
This is the initial phase shown in Figure \ref{fig:circuit_phases} (a).
It initializes the circuit when CLK1 is low, CLK2 is high and CLK3 is high leading to  $M5$ and $M6$ being ``ON'' and pre-charging the two output nodes to $V_{DD}$ such that $V_{out+} = V_{out-} = V_{DD}$. At the same time, $M9$ and $M10$ are ``OFF'' and the tail capacitors $C_{T}$ are discharged to GND through $M11$ and $M12$ resulting in $ V_{C_T}=0 $, where $V_{C_T}$ is the voltage across the tail capacitor.

\subsection{Amplification phase}
In the second phase, CLK1 turns high and CLK2 turns low while CLK3 is still high leaving all nodes at a high impedance state as shown in Figure \ref{fig:circuit_phases} (b). In this phase, the circuit acts like a differential amplifier typically as stated in section \ref{cs_paradigm}. The steady state of the differential output is proportional to the input voltage. The gain is mainly governed by $C_T$ and $C_{pL}$. Since there is no DC path to GND in this phase, the voltage difference is created by charge redistribution between the two output capacitors $C_{pL}$ and the tail capacitors $C_{T}$ while consuming little extra power consumption. At the end of this phase, the two output voltages are sitting at levels slightly higher than the voltage level $V_{min}$ as shown in Figure \ref{fig:time}. The output voltage difference is an amplified version of the input as explained before. The small-signal output voltage at the end  of the amplification phase can be given by: 

\begin{equation}
V_{out}= g_{m,avg} R_{out} V_{in}
\end{equation}

Where $g_{m,avg}$ is the average transconductance of the input pair and $R_{out}$ is the effective resistance of the switched capacitor structure resulting from charging and discharging of the load capacitors $C_{pL}$: $R_{out}=\frac{\Delta T}{C_{pL}} $. Where $\Delta T$ is the discharging time of $C_{pL}$ which equals the charging time of the tail capacitor $C_T$ to make $V_{p}$ rises from zero to $V_{CM}-V_{TH}-\Delta V $  where $\Delta V$ is somewhat small and arbitrary voltage.
$\Delta T$ is given in [\ref{paper:book_razavi}] as \begin{equation}  \Delta T=\frac{C_{T}}{K}\frac{V_{CM}-V_{TH}-\Delta V}{(V_{CM}-V_{TH})\Delta V}  \end{equation} where K is a physical constant of transistor parameters.

The output voltage at the end of the amplification phase can be given by: 
\begin{equation}
\label{equation:out_volt_amp}
V_{out}=2\times \frac{V_{CM}-V_{TH}-\Delta V}{V_{CM}-V_{TH}+\Delta V}\times \frac{C_T}{C_{pL}}\times V_{in} \end{equation}
\subsection{Regeneration phase}

The third phase shown in Figure \ref{fig:circuit_phases} (c) starts when CLK3 turns low. $M13$ turns ON, thus $M7$ and $M8$ form a cross-coupled regenerative latch with an initial condition given by the final value of the amplification phase. The differential output, hence, starts to grow exponential till it reaches its saturation limit where the upper limit is $V_{DD}$ and the lower limit is $V_{min}$. The cross-coupled 
devices work on regenerating the output voltage by charge redistribution between $C_{T}$, $C_{pL1}$ and $C_{pL2}$. Once one of the two outputs reaches $V_{DD}$ ($V_{out+}$), the regeneration operation stops and $M7$ turns ``OFF'' keeping ($V_{out-}$) sitting at a voltage $V_{min}$. The cross-coupled pair will prevent the lower output to be pulled high and hence preserves the regenerated output till the next reset phase.

\vspace{2ex}

On one hand, higher $C_{T}$ would lead to lower voltage levels after the charge-sharing operation taking place in the amplification phase. On the other hand, power consumption increases linearly with the $C_{T}$ value. In this design, a capacitance value of $30$ fF is chosen which resulted in an effective voltage gain at the end of amplification phase of 2.5 and a value of $V_{min}$ equal to 40\% of the supply.
This proposed architecture combines the pre-amplification and the regeneration functions of a typical comparator circuit into one stage. It also relies on the concept of charge redistribution and re-using the stored charges to obtain the required output voltage at the end of each phase. The comparator is sensitive to very small input voltages at extremely high speeds by enabling the regeneration after the amplification phase is done. The circuit consumes very low power consumption since there is no direct path from $V_{DD}$ to GND after the initial reset phase.

\vspace{2ex}

\begin{figure}[!t]
\centering
\includegraphics[height= 2.5in, width=2.1in]{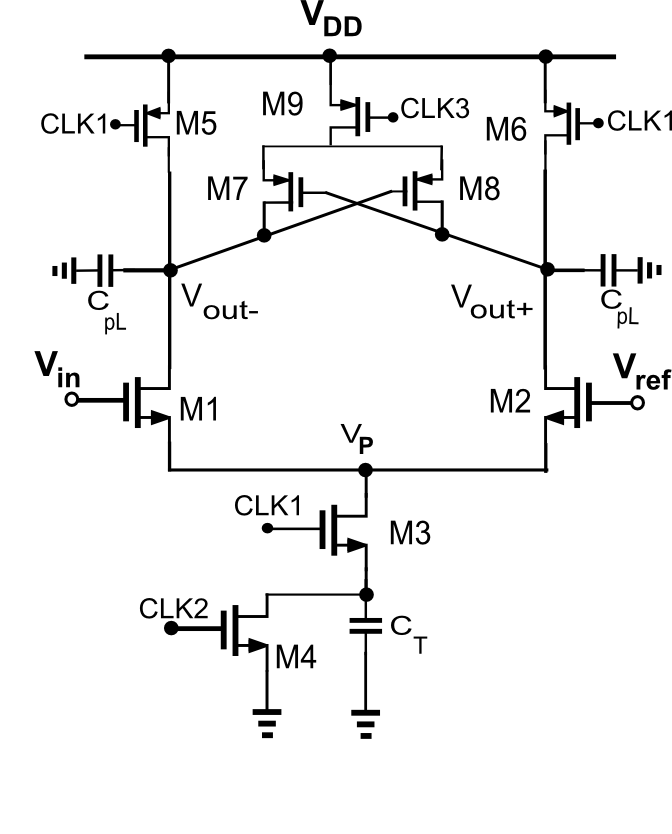}
\caption{A Single-Ended Version of the Proposed Comparator.}
\label{fig:single_end_comp}
\end{figure}

A single-ended version of the proposed comparator is shown in Figure \ref{fig:single_end_comp}. This single-ended circuit works the same as the differential circuit with same timing scheme. A single-ended version can be used with a single-ended input where a clocked comparator is needed and will be used for comparison with the conventional strongARM comparator.

%%%%%%%%%%%%%%%%%%%%%%%%%%%%%%%%%%%%%%%%%%%%%%%%%%%%%%%%%%%%%%%%%%%%%%%%%%%%%%%%%%%%%%%%%%%%%%%%%%%%%%%%%%%%%%%%%%%%%%%%%%%%%%%%%%%%%%%%%%%%%%%%%%%%%%%%%%%%%%%%%%%%%%%%%%%%%%%%%%%%%%%%%%%%%%%%%%%%%%%%%%%%%%%%%%%%%%%%%%%%%%%%%%%%%%%%%%%%%%%%%%%%%%%%%%%%%%%%%%%%%%%%%%%%%%%%%%%%%%%%%%%%%%%%%%%%%%%%%%%%%%%%%%%%%%%%%%%%%%%%%%%%%%%%%%%%%%%%%%%%%%%%%%%%%%%%%%%%%%%%%%%%%%%%%%%%%%%%%%
\subsection{Delay Analysis for the Proposed Circuit}
\label{sec:delay_proposed_diff_comp}

According to the work published in [\ref{ICM_paper_9}], the total delay for the conventional StrongARM comparator is comprised of two time delays, $\mathit{t}_{0}$ and $\mathit{t}_{latch}$. In our proposed circuit. The delay $\mathit{t}_{0}$ represents the time taken in amplification phase which is the quarter of the clock cycle $\mathit{T}_{s}$. The delay $\mathit{t}_{latch}$ is the time taken by cross-coupled latch to regenerate outputs to distinct values. It is defined as the time taken to have a voltage difference at the output ($\Delta{V}_{out}$) equal to $V_{DD}/2$ starting from an initial output voltage difference $\Delta{V}_{initial}$ from equation \ref{equation:out_volt_amp}. The total delay can be given as:

\begin{equation}
t_{delay}= t_{0}+t_{latch} 
\end{equation}

\begin{equation}
t_{0} = \frac{T_{s}}{4}
\end{equation}

\begin{equation}
t_{latch}= \frac{ \mathit{C}_{pL}} { \mathit{g}_{m,eff}} . ln (\frac{\Delta{V}_{out}}{\Delta{V}_{initial}}) = \frac{ \mathit{C}_{pL}} { \mathit{g}_{m,eff}} . ln (\frac{{V}_{DD}/2}{\Delta{V}_{initial}} )
\end{equation}

%\begin{equation}
%t_{latch}= \frac{ \mathit{C}_{pL}} { \mathit{g}_{m,eff}} .  ln (\frac{|\mathit{V}_{DD}|}{4 |\mathit{V}_{thp}| \Delta{V}_{in}} . \sqrt{ \frac{I_{tail}}{\beta_{1-4}}})
%\end{equation}

%\begin{equation}
%{I_{tail}= \frac{V_{CM}-V_{GS}} {R_{ON}} .exp(\frac{-t_{0}}{R_{ON} C_{T}})}
%\end{equation}

%$\beta_{1-4}$ is the input transistors current factor, $I_{tail}$ is a function of input common-mode voltage and $\mathit{VDD}$, and $\Delta{V}_{in}$ is the input difference voltage.

Where $\mathit{g}_{m,eff}$ is the effective transconductance of the back to back PMOS pair.

%%%%%%%%%%%%%%%%%%%%%%%%%%%%%%%%%%%%%%%%%%%%%%%%%%%%%%%%%%%%%%%%%%%%%%%%%%%%%%%%%%%%%%%%%%%%%%%%%%%%%%%%%%%%%%%%%%%%%%%%%%%%%%%%%%%%%%%%%%%%%%%%%%%%%%%%%%%%%%%%%%%%%%%%%%%%%%%%%%%%%%%%%%%%%%%%%%%%%%%%%%%%%%%%%%%%%%%%%%%%%%%%%%%%%%%%%%%%%%%%%%%%%%%%%%%%%%%%%%%%%%%%

\section{Charge-Steering Flash ADC}
\label{sec:flash_adc}

\begin{figure*}[!t]
\centering
\includegraphics[width=6in,height = 2.5in]{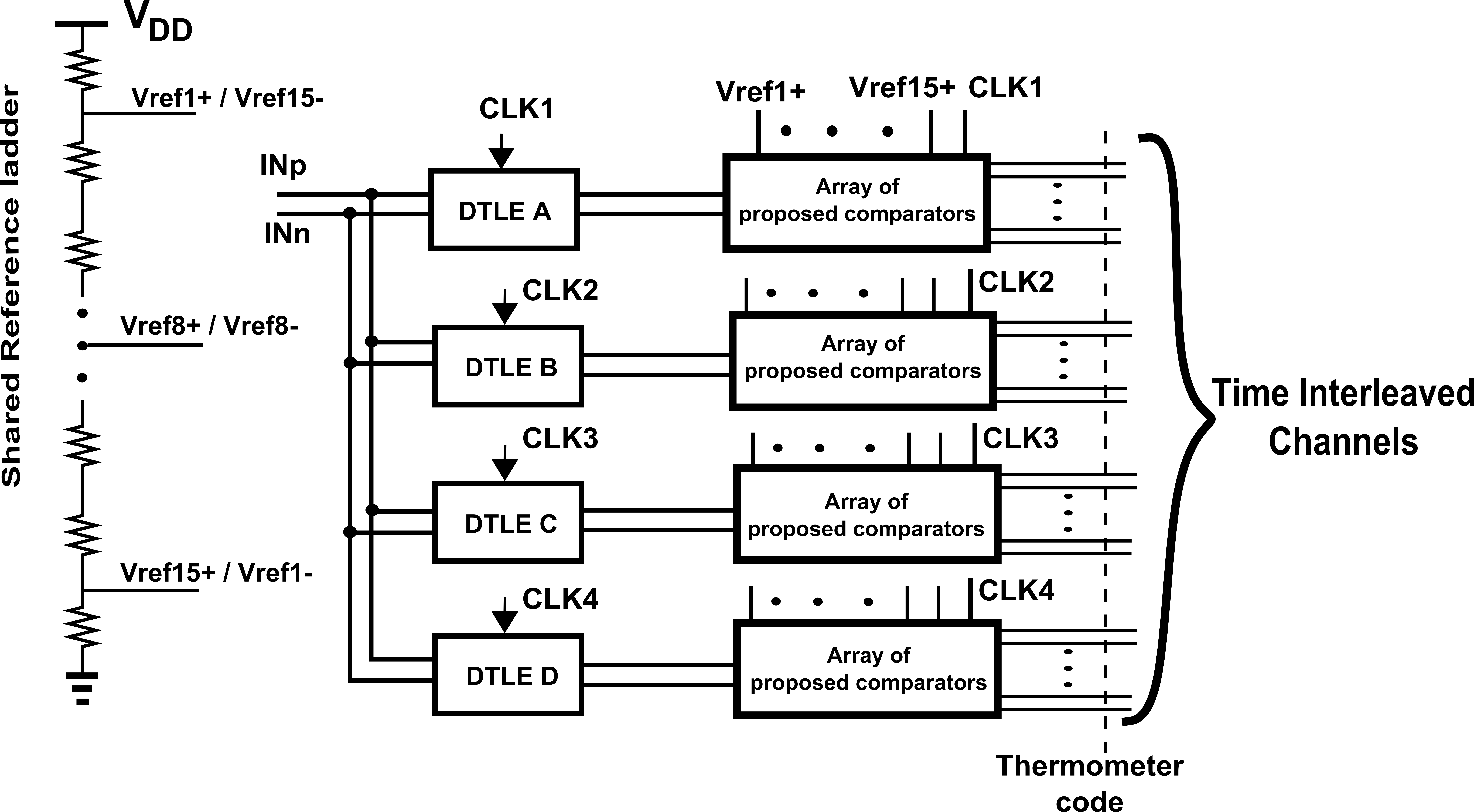}
\caption{Time-Interleaved ADC Architecture.}
\label{fig:adc_arch}
\end{figure*}

Figure \ref{fig:adc_arch} shows the flash ADC architecture. 4x interleaving is used while all four channels share the reference ladder to reduce the total power consumption. Four groups of 5 GHz clocks separated by $90^o$ phase shifts are used for the four channels. The proposed comparator is the main building block used in the ADC, it is followed by static RS latch to complete a flip-flop to successfully time the output with the clock edges. Sharing the reference ladder between the interleaved branches while keeping the kick-back noise at an acceptable level is also implemented and led to significant power reduction as well. The outputs of the four interleaved channels of Flash ADC were kept in thermometer code to be directly fed into digital signal processing (DSP) core for further digital feed-forward equalization (FFE) and/or decision-feedback equalization (DFE) according to the channel and link conditions. Moreover, This ADC can be used in a multi-standard system. It can be operated in a quarter-rate, a half-rate or a full-rate mode according to the required sampling rate. Noting that a DTLE circuit will replace the sample-and-hold circuit in each branch to equalize and sample and hold the input signal for each sub-ADC.

\subsection{Sample-and-Hold Network}
\label{SHA_theory}

Since the analog input is changing at very high speeds, Flash ADC-comparators might samples the input at different instances due to the different types of mismatches. To understand this problem, consider a low-frequency input is introduced to the ADC, the input will almost have a constant amplitude while the array of comparators sampling it to make proper decisions. But with the case of a high-frequency input signal, its fast-varying nature and the mismatches between comparators, the input might be sampled at different instances by the different comparators and this will give wrong decisions across the comparators array.

\vspace{2ex}

An optional sample-and-hold circuit was designed at the beginning to buffer between input nodes and the comparators, thus isolating the input from the high input capacitance from the comparators bank. The input devices were found to have a small input capacitance and since the proposed comparator is dynamic, it already samples the input at an instance of time, so the sampling circuit is eliminated. For now, input signal could be directly fed to and sampled by the dynamic comparators banks. Usually, in ADC-based equalizers, a linear equalizer precedes the ADC. Using a discrete-time linear equalizer can merge the functions of both blocks together as explained in section \ref{sec:dtle}.

\subsection{Clocking System and Buffers}

%\begin{figure}[!h]
%\centering
%\includegraphics[width=3in]{./needed_clks.png}
%\caption{The Required Clocks for All Branches.}
%\label{fig:all_clocks}
%\end{figure}

\begin{figure*}[!h]
\centering
\includegraphics[width=7in]{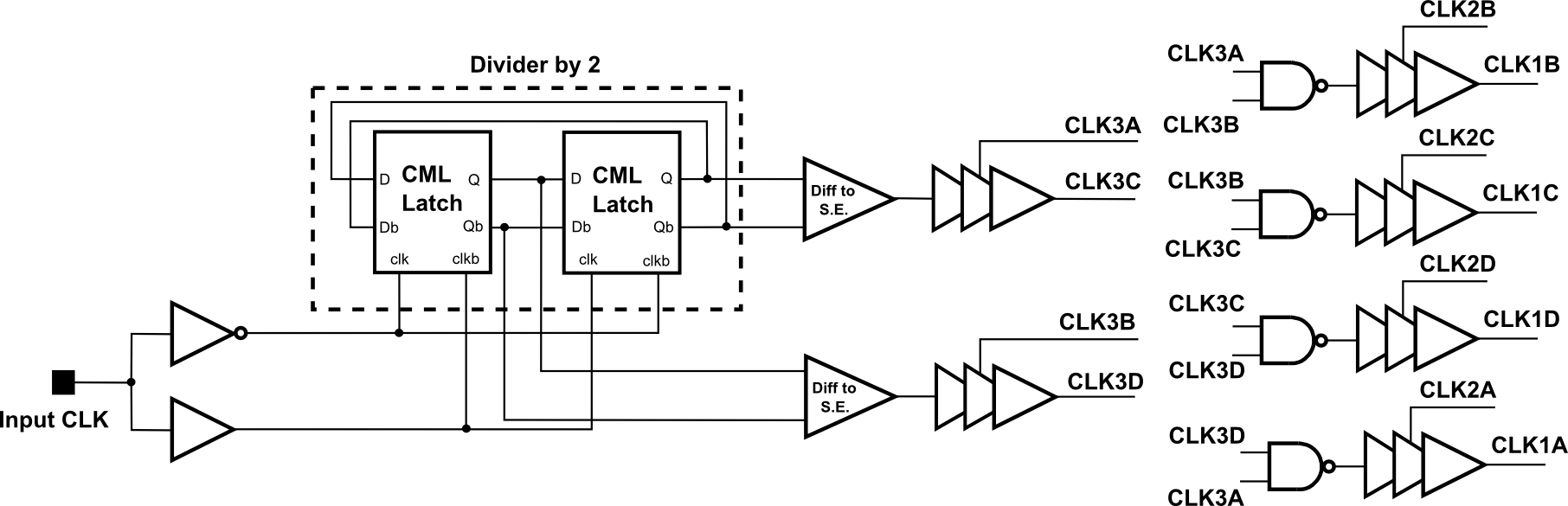}
\caption{The Designed Clocking System to Generate the Required Clocks.}
\label{fig:time_sys}
\end{figure*}

For the proposed ADC, we need 3 clocks for each channel as in Figure \ref{fig:time} and hence we need 4 groups of these 3 clocks, each group is shifted by $90^o$ from the other. To get these 12 clocks, there are two main thoughts, a quadrature PI or using logic operations between shifted clocks. The latter method is functional but quite inaccurate. However, we adopted it for testing and checking functionality.

\vspace{2ex}
Figure \ref{fig:time_sys} presents the designed timing system used to provide our ADC with required 12 different clocks. A 10-GHz clock is provided from a higher level, it is introduced as a single-ended clock, this signal is buffered and inverted then injected into 2 CML latches. These 2 CML latches form together, with the shown connection, a divide-by-2 circuit. Since CML latches have non-rail-to-rail outputs, the outputs are fed into differential to single-ended CML to CMOS converters. The generated clock signals are buffered and used through NAND gates and other buffers to generate the whole 12 different clock signals. Buffers are used to be able to drive the comparators.

\vspace{2ex}

CML latch circuit was used to work properly with the input 10-GHz clock in a 65-nm technology. The differential to the single-ended converter with buffers are working as a CML to CMOS converter. Conventional CMOS NAND gates were adopted to generate the $75\%$ duty cycle clocks. The buffer chain was used to generate the $25\%$ duty cycle clocks as well and to drive the loads, the switches inside the comparators.

%%%%%%%%%%%%%%%%%%%%%%%%%%%%%%%%%%%%%%%%%%%%%%%%%%%%%%%%%%%%%%%%%%%%%%%%%%%%%%%%%%%%%%%%%%%%%%%%%%%%%%%%%%%%%%%%%%%%%%%%%%%%%%%%%%%%%%%%%%%%%%%%%%%%%%%%%%%%%%%%%%%%%%%%%%%%%%%%%%%%%%%%%%%%%%%%%%%%%%%%%%%%%%%

\section{Simulation Results}
\label{sec:sims}

\begin{figure}[!h]
\centering
\includegraphics[width=3.5in,height=1.5in]{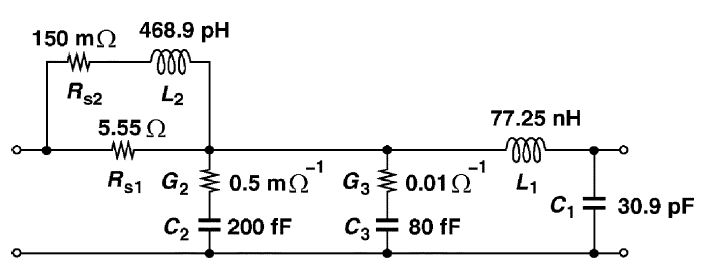}
\caption{A Schematic for Equivalent Model of 1-inch FR4 Channel. [\ref{paper:Gondi}]}
\label{fig:channel_1in}
\end{figure}

\begin{figure}[!h]
\centering
\includegraphics[width=3in]{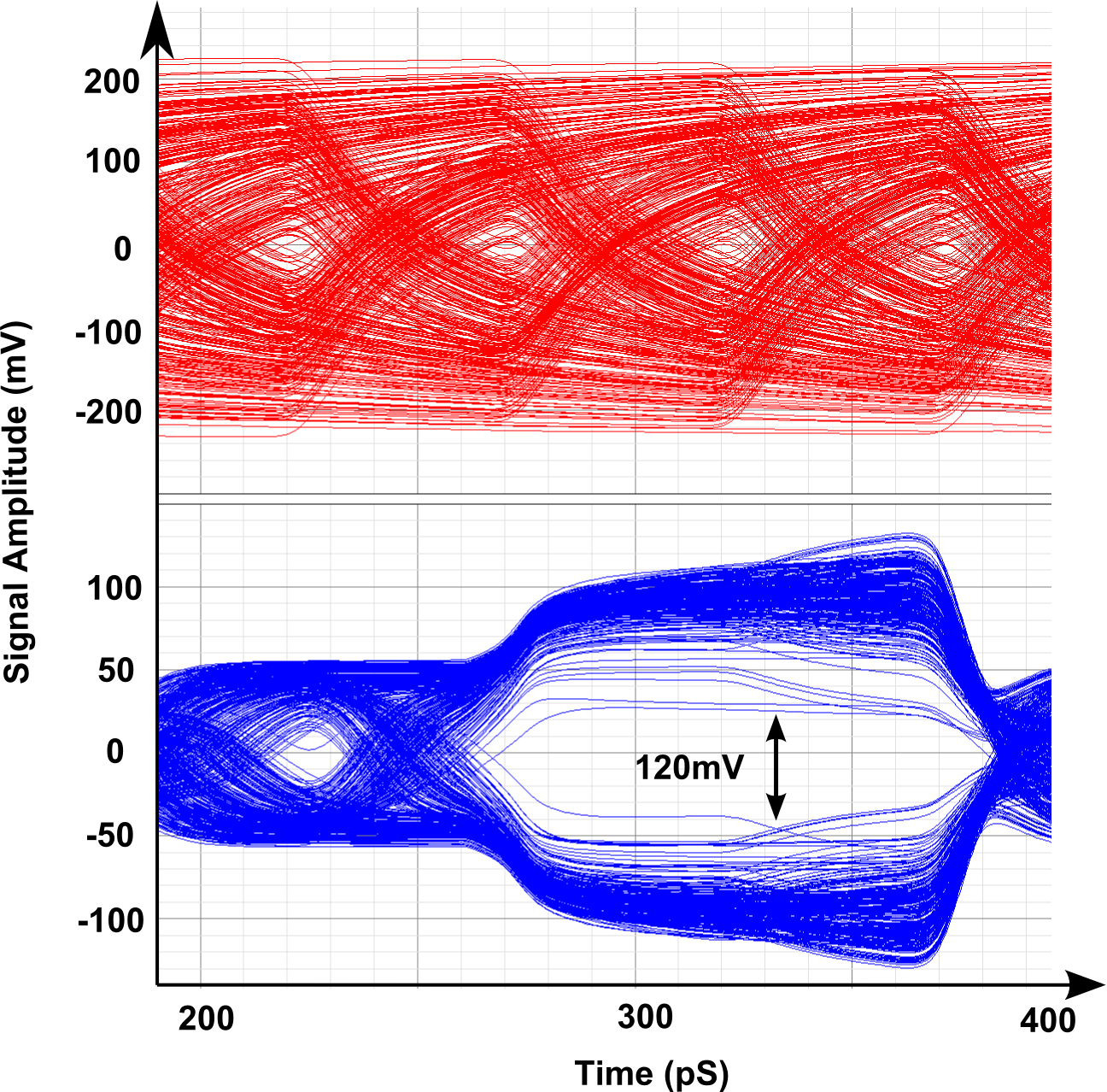}
\caption{The Eye Diagrams at the Receiver Input (Upper) and After DTLE.}
\label{fig:eye_diagrams}
\end{figure}

The DTLE was simulated across a 12-inch FR4 channel, the model of the channel used is given in Figure \ref{fig:channel_1in}. Data was transmitted through the channel with a 20-Gb/s rate and DTLE was running on 5-GHz clocks while consuming only 0.57 mW from a 1.2-V supply. The eye diagram at the end of the channel, i.e. input of the receiver, and the eye diagram after the DTLE are shown in Figure \ref{fig:eye_diagrams}. The first eye diagram is totally closed, while the latter eye diagram has a vertical eye opening of 120 mV. Since the ADC has a differential dynamic range of 600 $mV_{pp}$ and a common-mode input voltage of 750 mV, the DTLE needs to be followed by a programmable gain amplifier (PGA) to match the DTLE output with the dynamic range and the input common-mode voltage of the ADC. In summary, Table \ref{table:ctle_compare} compares the designed DTLE with the prior work in the literature.  

\begin{table*}[!h]
\centering
\renewcommand{\arraystretch}{1.3}
\caption{DTLE Performance Comparison.}
\label{table:ctle_compare}
\begin{tabular}{|l|c|c|c|c|}
\hline
Specification & [\ref{paper:Gondi}]  &[\ref{paper22}] & [\ref{paper23}] & This Work\\
\hline
Technology & 130nm & 32nm & 90nm & 65nm\\
\hline
Power Supply (V) & 1.6 & 1.1 & 1.25 & 1.2 \\
\hline
EQ. Architecture & Inductive-Loaded CTLE & CTLE + -ve C Bandwidth Extension & CTLE &  Quarter-Rate DTLE \\
\hline
Data Rate (Gb/s) & 10 & 12.5 & 8 & 5 (per channel) \\
\hline
Channel Loss (dB)  & 18 &  32 &N/A & 12\\
\hline
Gain @ Nyquist (dB) & N/A & 10 (Input Stage + 2-CTLE Stages) & N/A & 20\\
\hline
Power (mW) & 41 & 5.25 & 2.32 & 0.57*\\
\hline
\end{tabular}
\begin{flushleft} *for a single DTLE  \end{flushleft} 
\end{table*}

\begin{figure}[!h]
\centering
\includegraphics[height=2.2in, width=3.7in]{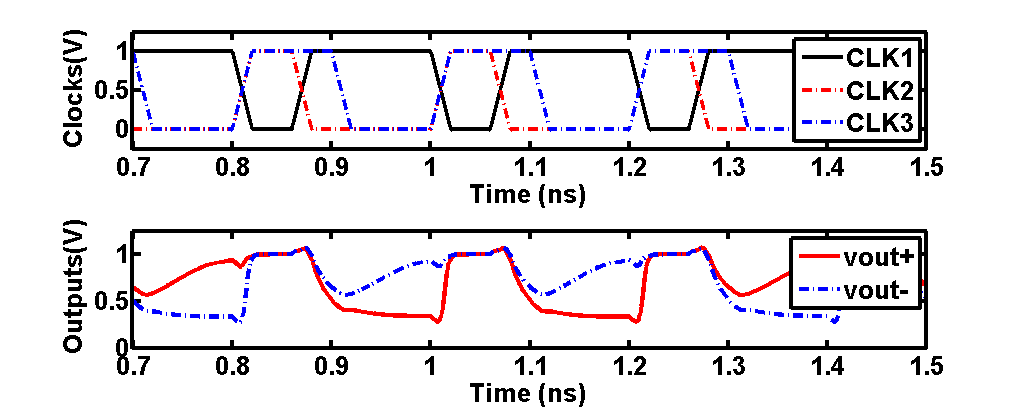}
\caption{Simulation Results for Comparator Outputs Vs. Clocking Scheme.}
\label{fig:result_1}
\end{figure}

\vspace{2ex}
The proposed 5-GHz comparator shown in Figure \ref{fig:proposed_comp_diff} has been designed using a digital 65-nm CMOS technology. The total power consumption of the proposed fully differential comparator is 189 $\mu$W and its single-ended version only consumes 66 $\mu$W. Simulated waveforms of clocks and outputs are shown in Figure \ref{fig:result_1}. The power consumption of the single-ended topology is split into 54 $\mu$W for the reset and amplification phases and 12 $\mu$W for the regeneration phase, whereas the differential comparator consumes 163 $\mu$W and 26 $\mu$W, respectively. In this comparator, Kick-back noise level is as tiny as 1 mV because outputs of the comparator core don't have rail to rail values.

%80\% of the input-referred offset is due to the input pair.  Monte-Carlo simulations show that the input-referred offset of the comparator is 9.55 mV while the input referred offset due to the input pair is 7.5mV. This shows that added circuitry results in only 20\% extra input-referred offset.

\vspace{2ex}
Monte-Carlo simulation for the offset %is shown in Figure \ref{fig:comp_montecarlo}. It 
shows that the input-referred offset sigma of the comparator is 14.4 mV. The input-referred offset sigma of the input devices only is found to be 13.9 mV contributing to 97\% of the total input-referred offset. This means that contribution of the newly added embedded regenerative latch is negligible when referred to the input. Keeping this in mind, making the input pair larger will drop down the input-referred offset significantly.

%\begin{figure}[!t]
%\centering
%\includegraphics[width=3in, height= 2in]{./montecarlo_sim1}
%\caption{Monte-Carlo Simulation Results for Offset. }
%\label{fig:comp_montecarlo}
%\end{figure}

\vspace{2ex}

A detailed comparison between the single-ended topology and a conventional StrongARM comparator is shown in Table \ref{table:comparison_comp}. The conventional StrongARM is known to be a low-power structure. For comparison purposes, the input pair dimensions are kept equal between the two designs.

\begin{table}[!h]
\centering
\renewcommand{\arraystretch}{1.3}
\caption{Comparison Between the Proposed Comparator and StrongARM Comparator at 5 GHz.}
\label{table:comparison_comp}
\begin{tabular}{|l|c|c|}
\hline
Comparison Point & StrongARM & This work\\
\hline
Power consumption ($\mu$W) & 256 & 66 (189 for diff-ended) \\
\hline
Max. speed for 5 mV input (GHz)& 5 & 6\\
\hline
Sensitivity at 5 GHz clock (mV)& 4 & 1\\
\hline
3-Sigma Offset(w/o the input pair) (mV) & 11.16 & 0.5\\
\hline
\end{tabular}
\end{table} 

\vspace{2ex}

\begin{figure}[!t]
\centering
\includegraphics[width=3in]{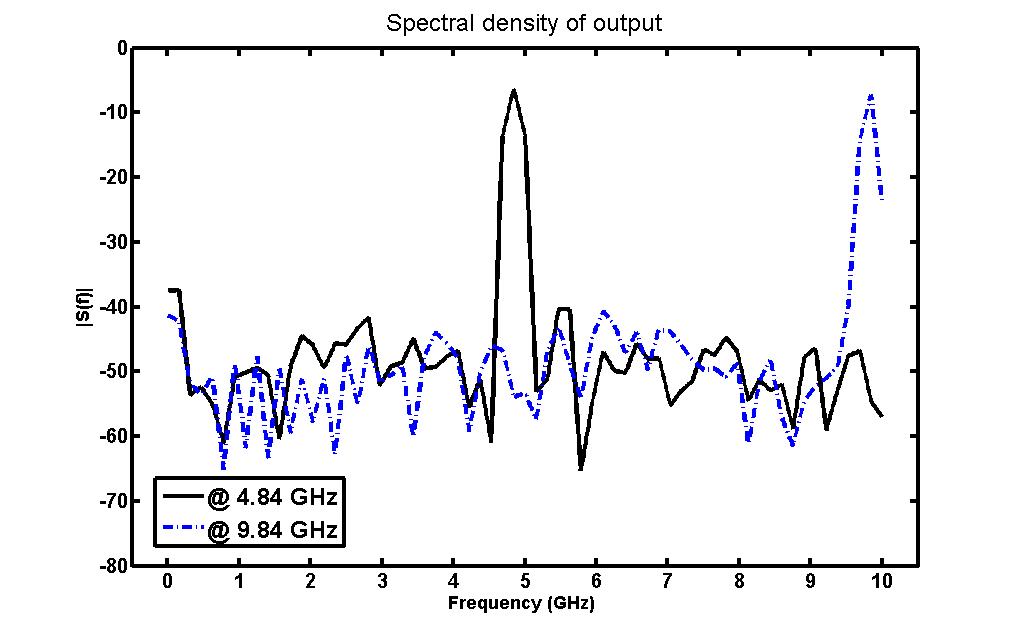}
\caption{Frequency Response of Output for Sinusoidal Inputs of 4.84 GHz and 9.84 GHz Input frequencies.}
\label{fig:fft}
\end{figure}

Clock buffers have been designed to drive switches and comparators. The capacitive loading per branch is 160 fF. The power consumption of clock buffers for each interleaved branch is 1 mW. This gives a total power consumption of 4 mW for the clock buffers of the whole ADC. The designed 4-bit 20-GS/s flash ADC has a total power consumption of $15.5$ mW.
A breakdown of the power consumption of different blocks is given in Table \ref{table:power break}. Figure \ref{fig:fft} shows the frequency response for the ADC output for sinusoidal inputs at 4.84 GHz and 9.84 GHz input frequencies while sampling at an effective frequency of 20 GHz. For the 9.84-GHz input frequency, Table \ref{table:all_branches} summarizes the performance of the whole ADC.

\begin{table}[!t]
\caption{Power Consumption Breakdown.}
\label{table:power break}
\centering
\begin{tabular}{|l||c|}
\hline
Ref. ladder & 200 $\mu$W \\
\hline
Comparators & 11.34 mW \\
\hline
Clock buffers & 4 mW \\
\hline
Total Power & 15.5 mW \\
\hline
\end{tabular}
\end{table} 

\begin{table}[!h]
%% increase table row spacing, adjust to taste
\renewcommand{\arraystretch}{1.3}
% if using array.sty, it might be a good idea to tweak the value of
%\extrarowheight as needed to properly center the text within the cells
\caption{The Designed ADC Performance Summary.}
\label{table:all_branches}
\centering
\begin{tabular}{|l||c|}
\hline
Sampling frequency & 20 GS/s\\
\hline
Input Range & 600 $mV_{diff}$\\
\hline
SFDR & 33.58 dB\\
\hline
SNDR & 23.86 dB\\
\hline
ENOB (for 9.84 GHz input frequency) & 3.67 Bits\\
\hline
Power at Nyquist frequency & 15.5 mW\\
\hline
FOMW*  & 60.8 fJ/conv-step \\
\hline
\end{tabular}
\begin{flushleft} 
* $\mathrm{FOMW=\frac{Power}{Sampling\, rate\,*2^{ENOB}}}$
\end{flushleft} 
\end{table} 

\vspace{2ex}

\begin{figure}[!t]
\centering
\includegraphics[width=3in, height=2in]{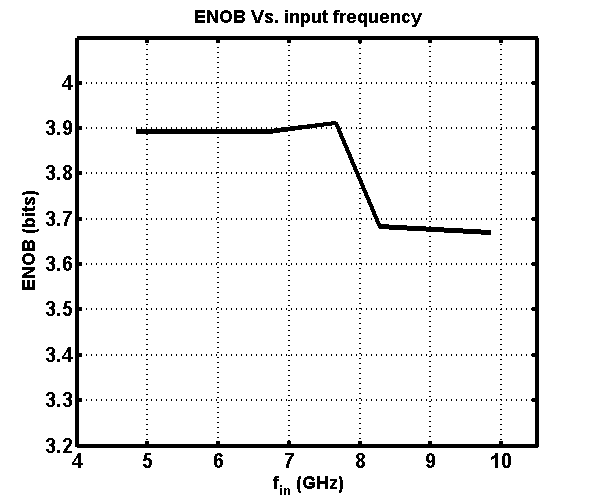}
\caption{ENOB Vs. Input Frequency.}
\label{fig:ENOB}
\end{figure}

Figure \ref{fig:ENOB} shows the resulting ENOB versus input frequency. It shows a peak ENOB of 3.9 bits at 7.66 GHz input signal.

%\begin{figure}[!t]
%\centering
%\includegraphics[width=3in,height=2in] {./SFDR_FREQ.jpg}
%\caption{SFDR Vs. Input Frequency.}
%\label{fig:SFDR}
%\end{figure}

\vspace{2ex}

\begin{figure}[!t]
\centering
\includegraphics[width=3in, height=1.5in]{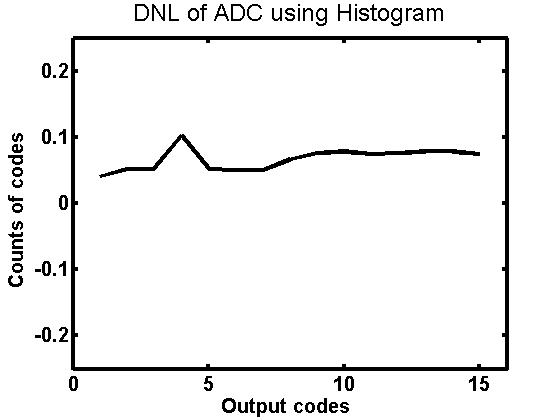}
\caption{DNL of the ADC.}
\label{fig:dnl}
\end{figure}

\begin{figure}[!h]
\centering
\includegraphics[width=3in, height=1.5in]{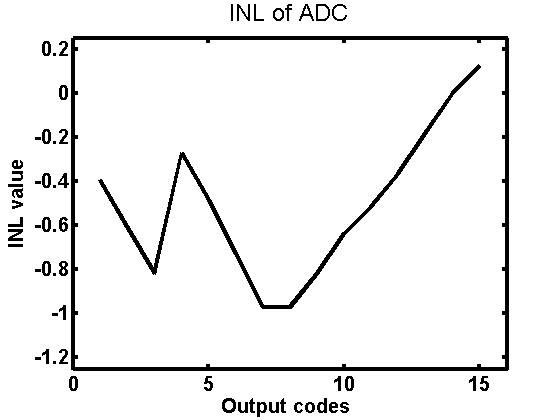}
\caption{INL of the ADC.}
\label{fig:inl}
\end{figure}

Small kick-back noise levels allow us to increase the used values of resistors in reference ladder, hence low power consumption for the ladder. DNL and INL for the ADC are shown in Figure \ref{fig:dnl} and Figure \ref{fig:inl} respectively.
 
\vspace{2ex}

Table \ref{table:comparison} shows a detailed comparison between this work and recently published ADCs, it shows that our proposed ADC is the best in class when compared to other high-speed ADCs.

\begin{table*}[!h]
\renewcommand{\arraystretch}{1.3}
\caption{Comparison Between this Work and Recently Reported Flash ADCs.}
\label{table:comparison}
\centering
\begin{tabular}{|l|c|c|c|c|c|c|c|c|}
\hline
Item & [\ref{paper12}] &[\ref{paper13}] & [\ref{paper14}] & This work \\
\hline
Process & 32-nm CMOS & 40-nm CMOS & 130-nm SiGe & 65-nm CMOS\\
\hline
Power (mW) & 69.5 &500 & 3000 & 15.5 \\
\hline
Bits & 6& N/A & 5& 4 \\
\hline
Sample rate (GS/s)& 20 & 25 & 22 & 20 \\
\hline
SNDR (dB) & 30.7 & 25.8 & 20 & 23.86 \\
\hline
FOMW* (fJ/conversion-step) & 172.3 & 1255.8 & 16695.8 & \textbf{ 60.8 }\\
\hline
\end{tabular}
\begin{flushleft} 
* $\mathrm{FOMW=\frac{Power}{Sampling\, rate\,*2^{ENOB}}}$
\end{flushleft} 
\end{table*} 

\begin{figure}[!h]
\centering
\includegraphics[width=3in]{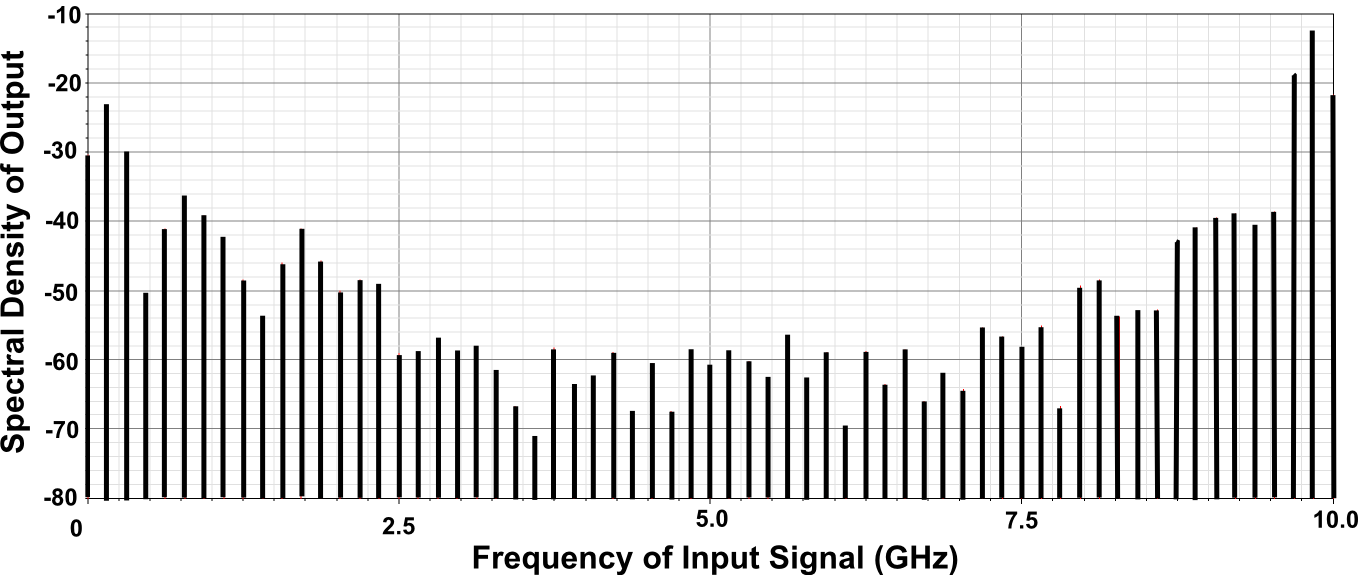}
\caption{The Spectral Density of the Extracted-ADC Output.}
\label{fig:RcCC_PEX_layout_ideal_clock}
\end{figure}

\begin{figure}[!h]
\centering
\includegraphics[width=3in]{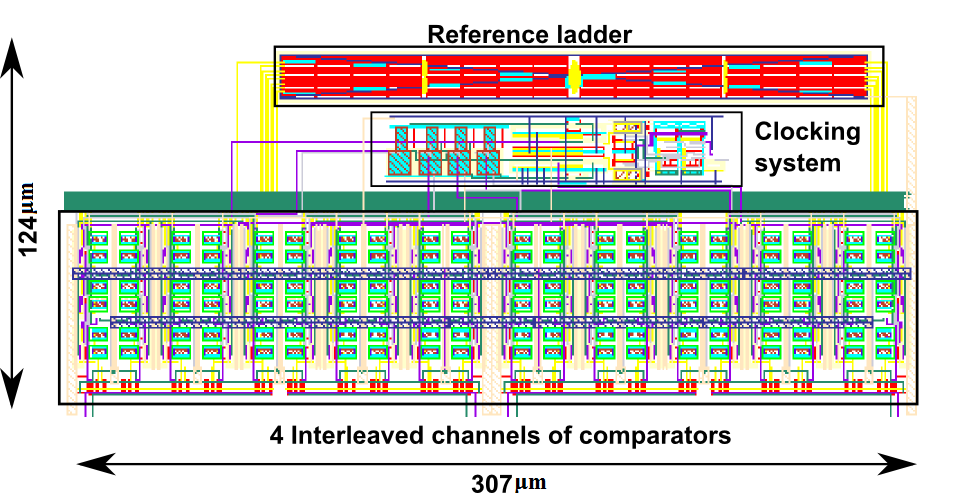}
\caption{Layout of the ADC.}
\label{fig:ADC_layout}
\end{figure}

\vspace{2ex}
Figure \ref{fig:RcCC_PEX_layout_ideal_clock} shows the FFT for the output of the whole ADC after parasitic extraction (PEX) working with the highest targeted speed of 20 GS/s. The Flash ADC with its clocking system and a testing current steering digital to analog converter (DAC) were laid-out in a total silicon area of 307$\mu$m*124$\mu$m. The layout is as shown in Figure \ref{fig:ADC_layout}. 

\section{Conclusion}
\label{sec:conc}

Charge-steering concept is an excellent candidate for high-speed circuits used in analog-mixed signal systems with low power requirements.
It proves an advantageous usage over the current-steering traditional concept, as well as the conventional StrongARM comparator architecture. Charge-steering is used in this paper to design an ultra-low power, 189-$\mu$W 5-GHz charge-steering comparator. The proposed idea uses an embedded regenerative latch which is only enabled after the amplification cycle is completed. The input referred offset of this extra latch is negligible compared to the input pair offset contribution. Furthermore, the proposed comparator is used to build a 4-bit 4x time-interleaved Flash ADC that works on 20 GS/s sampling rate. ADCs were used to be the bottleneck of the ADC-based receivers. With the ADC designed in this work, ADC's power consumption was as low as 15.5 mW, the thing that paves the way and relieves the burden facing the ADC in the high-speed designs. Moreover, the ADC was preceded by a modified version of a continuous-time linear equalizer, this modified version is called discrete-time linear equalizer (DTLE) and it merges a CTLE with a sample-and-hold network. This equalizer is used after a 12-inch FR4 channel to open the eye diagram of the received signal and enables it to be processed by the ADC and the digital equalizers in a DSP core after that.  

%\section{Acknowledgement}
%This work was supported by the Egyptian Information Technology Industry Development Agency (ITIDA) under the ITAC Program CFP No.79.


\begin{thebibliography}{}

\bibitem {1}
\label{paper:1}
S. Sheikhaei, S. Mirabbasi, and A. Ivanov, ``A 4-bit 5GS/s flash A/D converter in 0.18um CMOS,''  Proc. IEEE ISCAS, vol. 6, pp. 6138-6141, May 2005.

%\bibitem {}
%\label{ICM_paper_2}
%B. Goll and H. Zimmermann, ``A comparator with reduced delay time in 65-nm CMOS for supply voltages down to 0.65,'' IEEE Trans. Circuits Syst. II, Exp. Briefs, vol. 56, no. 11, pp. 810-814, Nov. 2009.

%\bibitem {}
%\label{ICM_paper_3}
%B. Goll and H. Zimmermann, ``A 0.12 $\mu$m CMOS comparator requiring 0.5V at 600MHz and 1.5V at 6 GHz,'' in Proc. IEEE Int. Solid-State Circuits Conf., Dig. Tech. Papers, pp. 316-317, Feb. 2007.


%\bibitem {}
%\label{ICM_paper_5}
%B. Goll and H. Zimmermann, ``A 65nm CMOS comparator with modified latch to achieve 7GHz/1.3mW at 1.2V and 700MHz/47$\mu$W at 0.6V,'' in Proc. IEEE Int. Solid-State Circuits Conf. Dig. Tech. Papers,  pp. 328-329, Feb. 2009.


%\bibitem {}
%\label{paper:strongArm_origin}
%J. Montanaro et al., ``A 160-MHz, 32-b, 0.5-W CMOS RISC microprocessor,'' IEEE Journal of Solid-State Circuits, vol. 31, no. 11, pp. 1703-1714, 1996.


%\bibitem {2}
%\label{paper2}
%M. Harwood et al., ``A 12.5 Gb/s SerDes in 65 nm CMOS using a baud-rate ADC with digital receiver equalization and clock recovery,'' ISSCC Dig. of Tech. Papers, pp. 436-437, Feb. 2007.

%\bibitem {3}
%\label{paper3}
%C.-K. K. Yang and E.-H. Chen, ``ADC-based serial I/O receivers,''  Proc. Custom Integr. Circuits Conf., pp. 323-330, Sep. 2009.

%\bibitem {4}
%\label{paper4}
%H. Chung and G.-Y. Wei, ``Design-space exploration of backplane receivers with high-speed ADCs and digital equalization,''  Proc. Custom Integr. Circuits Conf., pp. 555-558, Sep. 2009.

%\bibitem {5}
%\label{paper5}
%H. Chung, ``A 7.5-GS/s 3.8-ENOB 52-mW flash ADC with clock duty cycle control in 65nm CMOS,'' VLSIC, pp. 268-269, June 2009.

%\bibitem {6}
%\label{paper6}
%E.-H. Chen and Chih-Kong Ken Yang, ``10Gb/s Serial I/O Receiver Based on Variable Reference ADC,'' VLSIC, pp. 288-289, June 2011.

%\bibitem {7}
%\label{paper:murmann}
%http://web.stanford.edu/~murmann/adcsurvey.html

%\bibitem {8}
%%\label{paper12}
%Chen, V.H.-C.; Pileggi, L., ``A 69.5mW 20GS/s 6b time-interleaved ADC with embedded time-to-digital calibration in 32nm CMOS SOI,'' ISSCC Dig. of Tech. Papers, pp.380,381, 9-13 Feb. 2014.

\bibitem {}
\label{paper:Ismail_DTLE}
A. Ismail, S. Ibrahim and M. Dessouky, ``An 8Gbps discrete time linear equalizer in 40nm CMOS technology,'' IEEE 58th International Midwest Symposium on Circuits and Systems, Fort Collins, CO, pp. 1-4, 2015.

\bibitem {}    
\label{paper:Gondi}
S. Gondi and B. Razavi, ``Equalization and Clock and Data Recovery Techniques for 10-Gb/s CMOS Serial-Link Receivers,'' IEEE Journal of Solid-State Circuits, vol. 42, no. 9, pp. 1999-2011, 2007.



\bibitem {10}
\label{paper:CS_razavi}
B. Razavi, ``Charge steering: A low-power design paradigm,'' Custom Integrated Circuits Conf., pp. 1-8, Sep. 2013.

\bibitem {11}
\label{paper:CS_Feb2015}
Jun Won Jung, Razavi, B., ``A 25 Gb/s 5.8 mW CMOS Equalizer,'' IEEE Journal of Solid-State Circuits, vol.50, no.2, pp. 515,526, Feb. 2015


%\bibitem {}
%\label{paper:ayesh_ICM}
%M. M. Ayesh, S. Ibrahim and M. M. Aboudina, ``Design and analysis of a low-power high-speed charge-steering based StrongARM comparator,'' International Conference on Microelectronics, 2016, pp. 209-212.


\bibitem {9}
\label{paper:embedded_latch}
Yun-Ti Wang; Razavi, B., ``An 8-bit 150-MHz CMOS A/D converter,'' IEEE Journal of Solid-State Circuits, vol.35, no.3, pp.308,317, March 2000.



\bibitem {}
\label{ICM_paper_15}
M. M. Ayesh, S. Ibrahim and M. M. Aboudina, ``A 15.5-mW 20-GSps 4-bit charge-steering flash ADC,'' IEEE International Midwest Symposium on Circuits and Systems, Fort Collins, CO, pp. 1-4, 2015.



\bibitem {}
\label{ICM_paper_9}
S. Babayan-Mashhadi and R. Lotfi, ``Analysis and Design of a Low-Voltage Low-Power Double-Tail Comparator,'' in IEEE Transactions on Very Large Scale Integration (VLSI) Systems, vol. 22, no. 2, pp. 343-352, Feb. 2014.


\bibitem {12}
\label{paper:book_razavi}
B. Razavi, ``Design of Analog CMOS Integrated Circuits,'' NY, USA: McGraw-Hill, 2001.


\bibitem {67}
\label{paper22}
T. Toifl et al., ``A 2.6 mW/Gbps 12.5 Gbps RX with 8-tap switched-cap DFE in 32 nm CMOS,'' IEEE Symposium on VLSI Circuits, pp. 210–211, 2011.

\bibitem {68}
\label{paper23}
Young-Sik Kim et.al., ``An 8Gb/s quad-skew-cancelling parallel transceiver in 90nm cmos for high speed dram interface,'' in IEEE International Solid-State Circuits Conference ISSCC, Feb 2012, pp. 136 – 138.

\bibitem {8}
\label{paper12}
Chen, V.H.-C.; Pileggi, L., ``A 69.5mW 20GS/s 6b time-interleaved ADC with embedded time-to-digital calibration in 32nm CMOS SOI,'' ISSCC Dig. of Tech. Papers, pp.380,381, 9-13 Feb. 2014.

\bibitem {13}
\label{paper13}
Crivelli, D., et al. ``A 40nm CMOS single-chip 50Gb/s DP-QPSK/BPSK transceiver with electronic dispersion compensation for coherent optical channels,'' ISSCC Dig. of Tech. Papers, pp.328-330, 6-9 Feb. 2012.

\bibitem {14}
\label{paper14}
Schvan, P.; Pollex, D.; Shing-Chi Wang; Falt, C.; Ben-Hamida, N., ``A 22GS/s 5b ADC in 130nm SiGe BiCMOS,'' ISSCC Dig. of Tech. Papers, pp.2340-2349, 6-9 Feb. 2006.




\end{thebibliography}
\end{document}